%% file: lowresource_tts.tex
\documentclass[conference]{IEEEtran}
\IEEEoverridecommandlockouts

\pdfoutput=1

\usepackage{amsmath,amssymb,amsfonts}
\usepackage{graphicx}
\usepackage{epstopdf}
\usepackage[sort]{cite}
\usepackage{lipsum}

\usepackage{scrextend}

\usepackage{subcaption}
\usepackage{booktabs}

\usepackage{tikz}
\usepackage{standalone}
\usepackage[hidelinks]{hyperref}
\usepackage{balance}

\usepackage[utf8]{inputenc}
\usepackage{pgfplots}
\usepackage{bm}
\DeclareUnicodeCharacter{2212}{−}
\usepgfplotslibrary{groupplots,dateplot}
\usetikzlibrary{patterns,shapes.arrows}
\pgfplotsset{compat=newest}

\usetikzlibrary{positioning,fit,calc, shapes}
\tikzstyle{block} = [draw=none, thick, text width=1.4cm, minimum height=0.35cm, align=center, fill=blue!25]  
\tikzstyle{oper} = [draw=none, ellipse, thick, text width=1.1cm, minimum height=0.1cm, align=center, fill=red!15]
\tikzstyle{arrow} = [thick,->,>=stealth]
\tikzset{
  font={\fontsize{9pt}{10}\selectfont}}
  
\title{Low-Resource Text-to-Speech Using Specific Data and Noise Augmentation}

\author{
    \IEEEauthorblockN{Kishor Kayyar Lakshminarayana\IEEEauthorrefmark{1}, Christian Dittmar\IEEEauthorrefmark{1}, Nicola Pia\IEEEauthorrefmark{1}, Emanu\"{e}l Habets\IEEEauthorrefmark{2}}
    \IEEEauthorblockA{\IEEEauthorrefmark{1}Fraunhofer Institute for Integrated Circuits (IIS), Erlangen, Germany
    }
    \IEEEauthorblockA{\IEEEauthorrefmark{2}International Audio Laboratories Erlangen, Germany \thanks{$^\dag$A joint institution of the Friedrich-Alexander-Universit{\"a}t Erlangen-N{\"u}rnberg (FAU) and Fraunhofer Institute for Integrated Circuits (IIS).}
    }
}

\def\baselinestretch{0.96}

\begin{document}
\maketitle
\input{abstract}
\input{intro}
\input{method}

\input{eval}
\input{discuss}
\input{ack}
\input{lastone}

\bibliographystyle{IEEEbib}

\bibliography{papers}

\end{document}

%% file: abstract.tex
\pdfoutput=1
\begin{abstract}
Many neural text-to-speech architectures can synthesize nearly natural speech from text inputs. These architectures must be trained with tens of hours of annotated and high-quality speech data. Compiling such large databases for every new voice requires a lot of time and effort. In this paper, we describe a method to extend the popular Tacotron-2 architecture and its training with data augmentation to enable single-speaker synthesis using a limited amount of specific training data. In contrast to elaborate augmentation methods proposed in the literature, we use simple stationary noises for data augmentation. Our extension is easy to implement and adds almost no computational overhead during training and inference. Using only two hours of training data, our approach was rated by human listeners to be on par with the baseline Tacotron-2 trained with 23.5 hours of LJSpeech data. In addition, we tested our model with a semantically unpredictable sentences test, which showed that both models exhibit similar intelligibility levels. 
\end{abstract}

%% file: intro.tex
\pdfoutput=1
\section{Introduction}
\label{sec:intro}

Modern neural text-to-speech (TTS) architectures such as Tacotron-1 and 2 \cite{gibiansky2017deep, shen_natural_2018} require large quantities of paired text and high-quality speech recordings for each speaker \cite{xu2020lrspeech} to synthesize near-natural speech from text input. For example, the popular LJ Speech \cite{ljspeech17} database, used extensively by the TTS community, has 23.5 hours of single-speaker samples. Collection and annotation of such a large database is resource-intensive, cumbersome, and expensive. Hence, there is a need for a TTS model that can be trained with a limited amount of paired text and speech data. 

For well-resourced languages like English, there are a few publicly available TTS databases with samples from multiple speakers. Hence, transfer learning with these databases often complements training with a low-resource speaker like in \cite{gibiansky2017deep, tits2019exploring}. It is shown in \cite{xu2020lrspeech} that transfer learning also works across languages. However using transfer learning could result in an unintentional transfer of speaking style or accent to the target speaker \cite{latorre21_ssw}.  

Data augmentation has been successfully used in neural network applications such as automatic speech recognition \cite{Park2019} and speaker verification \cite{Wu2019}. Recently, multiple data augmentation techniques have been proposed for TTS as well.  For example, \cite{shah_non-autoregressive_2021, huybrechts2021low} use the CopyCat \cite{karlapati20_interspeech} voice conversion (VC) model to create augmentation samples. Articles like \cite{sharma20b_interspeech, hwang2021tts, song22d_interspeech} use a teacher TTS to generate augmentation samples. This means that both the VC and TTS augmentation methods require a different neural architecture to be trained to produce the augmentation samples. Alternatively, data augmentation by changing the pitch of recorded speech was explored in \cite{cooper20_interspeech, 9616305}. These were multi-speaker TTS scenarios using large training data sets of the order of five hours per speaker. Alternatively, \cite{chung2019semi} used unpaired speech and text data for augmentation, the collection of which is also time consuming. A recent article \cite{9746291} explores data augmentation using re-ordering of the text-speech pairs, which requires meticulous preprocessing. 

This paper proposes a method for single-speaker low-resource TTS training using specific data and noise augmentation. To prevent the noise augmentations from degrading the output quality, we extend the Tacotron-2 \cite{shen_natural_2018} architecture with additional augmentation embeddings. Although Tacotron is not the latest TTS architecture, it remains a competitive baseline in many recent studies, e.g., \cite{song22d_interspeech, 9746291}. In contrast to existing transfer learning approaches, the proposed approach does not require a pre-trained model. We also provide the specifics of the training data, which can be applied to any new voice or language to reduce the time and effort to collect the data.

Listening tests show that our approach achieves statistically equivalent Mean Opinion Scores (MOS) in comparison to a baseline Tacotron-2 model trained with the complete 23.5 hours of LJ-Speech \cite{ljspeech17} data, while ours only uses a 2-hours subset of the same corpus. We further demonstrate that the two approaches generate similar text and speech alignment, which is a critical requirement for synthesis. Additionally, we verified our proposed approach with objective intelligibility metrics. 

%% file: method.tex
\pdfoutput=1
\section{Proposed Method}
\label{sec:proposed}

\begin{figure}[t]
  \centering
  \includegraphics[width=\linewidth]{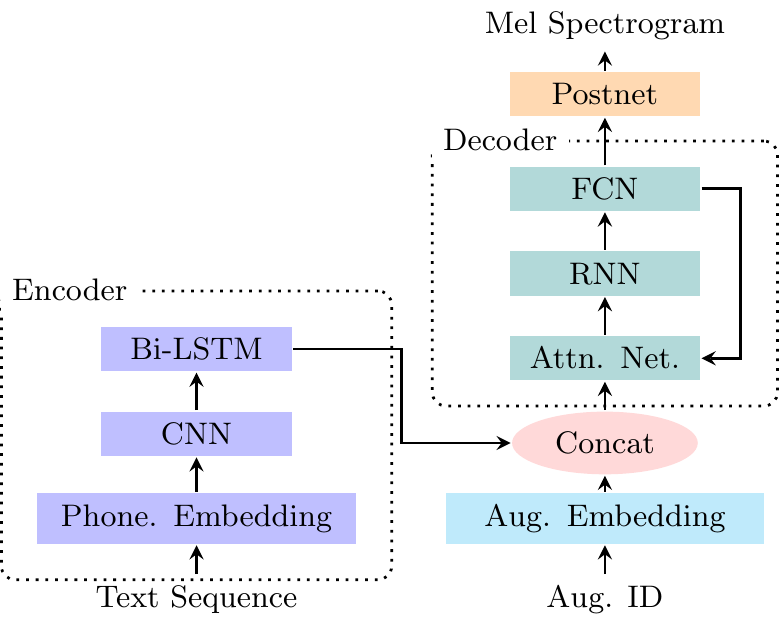}
  \caption{Block diagram of Tacotron-2 with proposed noise augmentation embedding. Here, Bi-LSTM is Bidirectional Long Short Term Memory, CNN is convolutional neural network (NN), RNN is recurrent NN and, FCN is Fully Connected Network. }
  \label{fig:proposed_method}
  \vspace{-1em}
\end{figure}

We use the popular auto-regressive Tacotron-2 \cite{shen_natural_2018} to implement the low-resource TTS. The architecture with the proposed augmentation embedding extension is shown in Fig.\ref{fig:proposed_method}. \mbox{Tacotron-2} converts an input text sequence to an output mel-spectrogram (mel) sequence via a sequence-to-sequence architecture. 

Tacotron-2 uses an attention model to learn the alignment between each input text token and the corresponding mel frames. Typically, this attention model gives higher weights to the input at the current time. Hence, these attention weights indicate the alignment of the input text and output mel frames. Critically, Tacotron-2 cannot synthesize intelligible speech if it fails to learn proper alignment. Smaller training datasets, in a low-resource case, often result in memorization. i.e., only the text sequences used for training can be synthesized, while unseen textual inputs lead to garbled speech output. 

The following sections describe our proposed methods and extensions to enable better alignment learning with limited training data.

\subsection{Training Data Specification}
\label{subsec:data_spec}

Like all neural networks, the training process of the \mbox{Tacotron-2} model is carried out in batches. This means that the update of the weights happens once per batch by averaging the loss and deriving the corresponding weight update. In a typical training session, the samples used in a batch are randomly chosen, resulting in a wide range of sample durations. Additionally, this batch processing requires zero padding, which is expected to be discarded by the attention model. The attention model gets a single update over these varying durations. For a low-resource database, this effect and the small number of available batches result in poor alignment between text and mels. Though reduced batch size may reduce the duration variation, it makes the gradient noisier and causes memorization. 

In this paper, we propose to reduce the variation within the training batches by providing a specification for the low-resource training data of the order of two hours. We recommend keeping the speech sample durations in the training dataset as close to one another as possible. For this purpose, we could use sentences or sentence segments (clauses) of similar length as the training data. A preprocessing step of splitting long sentences into short segments and otherwise using short sentences would achieve this. Further, this duration specification criterion should be used with the default consideration of having a broad, balanced phoneme distribution across the training set. 

These similar-length training samples help the attention network learn better alignment between the text and speech. Moreover, the need for zero padding becomes minimal, offering an easier learning task for the attention network. Further, the ability of the model to synthesize longer sentences is not compromised since the component phoneme durations are not affected by the sentence being long. As a tradeoff, the prosodic interdependencies across long sentences are not learned using the current approach due to the use of short segments. 

\subsection{Noise Augmentation}
\label{subsec:noise_aug}

We propose using noise augmentation to increase the number of samples available for training in a low-resource setting. Such an augmentation can create multiple samples with the same phoneme/speech content but significantly different mels. Using such noise-augmented samples can lead to degradation of quality, which is then avoided using the augmentation labels.  

A speaker label is often provided along with the text as an input to TTS models to synthesize different voices, e.g., \cite{gibiansky2017deep, jia2018transfer}. These labels help to generate neural embeddings as speaker representations. Similarly, we propose neural augmentation embeddings using specified augmentation identifiers (Aug. ID.). These are concatenated to the output of the Tacotron-2 encoder as shown in Fig. \ref{fig:proposed_method}. These neural augmentation embeddings are the common representation (multi-dimensional vector) learned across the samples of the distinct augmentation sets. For every augmentation set, a stationary noise with a known statistical distribution at a specific Signal-to-Noise Ratio (SNR) is added to each training sample. We used three kinds of stationary noise types, empirically defined, each at different SNRs. The three noise types were assigned to three Aug. IDs. The original data is associated with the "clean" Aug. ID, later used for inference. We found that using three different noise types for augmentation resulted in the same outcome, regardless of the specific noise types used. We also found that adding more than three noise augmentations would increase the training time without significant performance gain.

Intuitively, the augmentation embedding layer learns the common properties of the noise (or no) augmentation. Hence the rest of the model parameters are influenced by the relation between the input phonemes and the corresponding mels.

%% file: eval.tex
\pdfoutput=1
\section{Experimental Setup}
\label{sec:exp}

\subsection{Datasets}
\label{subsec:datasets}

We simulated the training data specification proposed in Section \ref{subsec:data_spec} using subsets totaling 2 hours from the LJ Speech database. Through experimentation, we discovered that using less than 2 hours of data does not produce usable speech, even with noise augmentation. As a baseline, we randomly selected 2 hours of data from the prompts resulting in a 2H\_RS (random set) dataset. Then, we arranged the whole 23.5-hour dataset in the order of durations and picked sentences (starting with the shortest duration) totaling 2 hours, resulting in 2H\_IS (informed set) dataset. The arrangement ensured that the samples were of short duration and that the durations in a training batch were close. We then used both 2H\_RS and 2H\_IS datasets with and without noise augmentation. 

We used white Gaussian noise (WGN) at 25~dB SNR, United States of America Standards Institute (USASI) standard noise at 15~dB SNR, and noise simulated from the noise power spectral density of a Knowles EM-3346 electret microphone (sensor noise) at 20~dB SNR as the three additive noises \cite{brookes_vbox_nodate}. We used the Voicebox toolbox \cite{brookes_vbox_nodate} for the noise addition. This determines the SNR based on the active speech power following the ITU-T P.56 recommendation. We found that specific noise types and SNR levels did not affect the results as long as there were three different augmentations.   

\subsection{Neural Vocoder}
\label{subsec:neural_vocoder}

We used a pre-trained StyleMelGAN \cite{mustafa2021stylemelgan} as the neural vocoder for converting the predicted mels to speech. The vocoder was trained for German synthesis with PAVOQUE \cite{steiner2013pavoque}, CSS\_10 \cite{park2019css10}, BITS \cite{ellbogen-etal-2004-bits}, and a proprietary two-speaker (one male and one female) dataset. This vocoder achieved good quality in listening tests conducted for English speech with copy synthesis and hence was used for the current evaluation. As an important side note, the vocoder can be trained with unlabelled speech data which is easier to collect and more readily available.

\subsection{Experiments}
\label{subsec:low-res-exp}

We simulated a low-resource scenario using LJ Speech data. We used the Tacotron-2 model trained with the full 23.5~hour LJ Speech dataset as the baseline. We then used the two separate low-resource datasets mentioned in Sec. \ref{subsec:datasets} for training. These are the 2H\_RS, a randomly selected low-resource dataset and the 2H\_IS, which simulates the proposed training data specification. These datasets were used to train the baseline Tacotron-2 model resulting in 2H\_RS and 2H\_IS models, respectively. We further trained the proposed model shown in Fig.\ref{fig:proposed_method} using these datasets and their noise-augmented versions, separately for RS and IS sets, resulting in 2H\_RS\_NA and 2H\_IS\_NA models. The training was done for 310K iterations, each with a batch size 24. Audio samples are shared at https://s.fhg.de/lrtts. 

Since the results from a single dataset training could be attributed to chance, we further investigated low-resource speakers `bdl' (male) and `slt' (female) from the CMU ARCTIC English speech synthesis database \cite{kominek2004cmu}. The results here were comparable. i.e., we could synthesize intelligible and good quality speech from `bdl'/`stl' samples only after using specific samples by excluding ``long" samples and using augmentation. All these experiments were different single-speaker training sessions.

\section{Results}
\label{sec:results}

\subsection{Text to Speech Alignment Learning}
\label{subsec:align_learn}

\begin{figure}
     \centering
     \begin{subfigure}[b]{\linewidth}
         \centering
         \includegraphics[width=\textwidth]{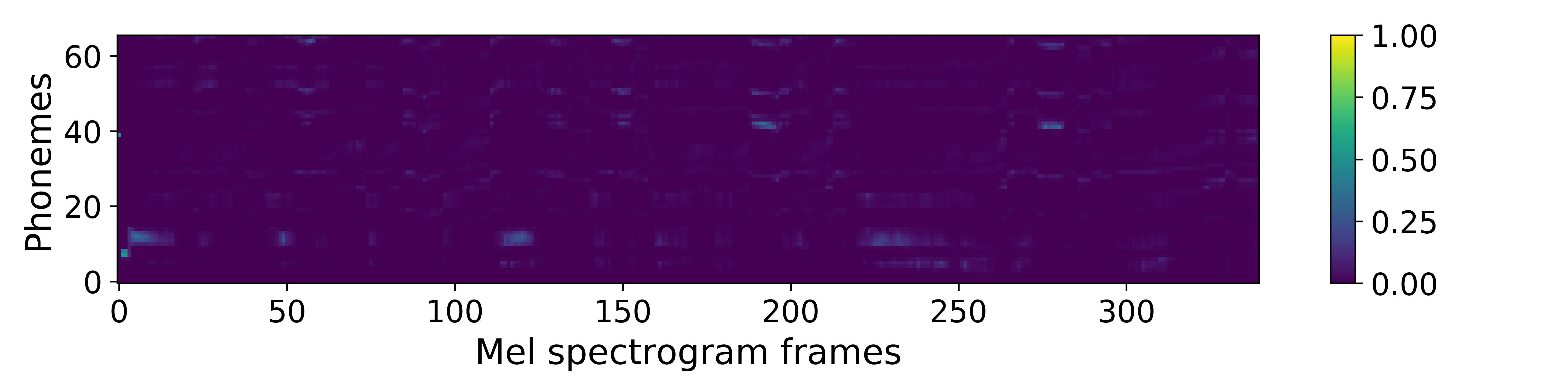}
         \caption{Alignment is not learned \textbf{without} noise augmentation with random training set (2H\_RS). }
         \label{fig:align}
     \end{subfigure}
     \hfill
     \begin{subfigure}[b]{\linewidth}
         \centering
         \includegraphics[width=\textwidth]{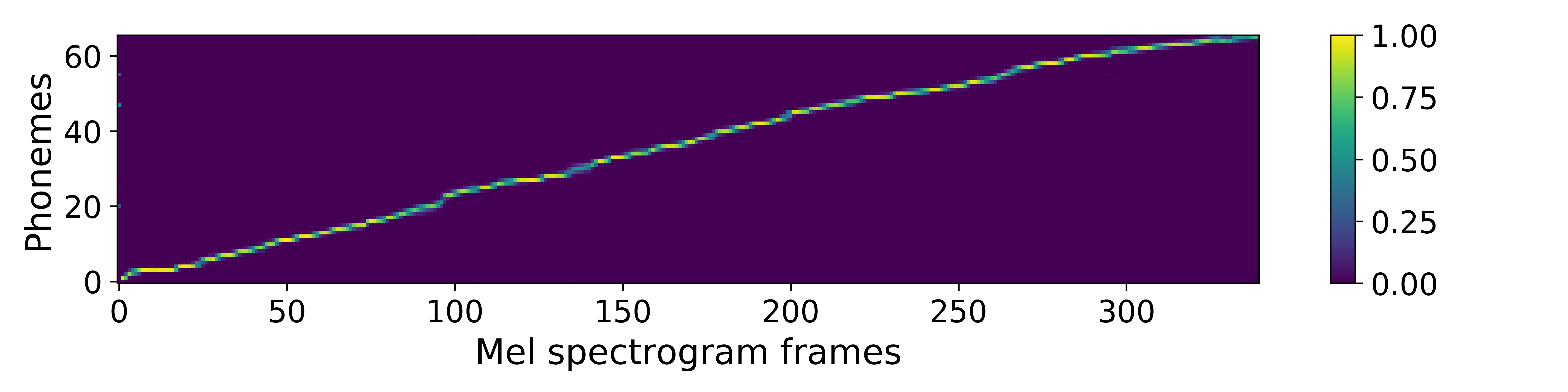}
         \caption{Alignment is learned \textbf{with} noise augmentation and informed training set (2H\_IS\_NA).}
         \label{fig:noise_align}
     \end{subfigure}
     \caption{Learned attention weights for LJ002-0114.wav from the LJ Speech dataset in two training sessions.}
     \label{fig:alignment}
\end{figure}
\vspace{0pt}
\begin{figure}[t]
  \centering
  \includegraphics[width=0.85\linewidth]{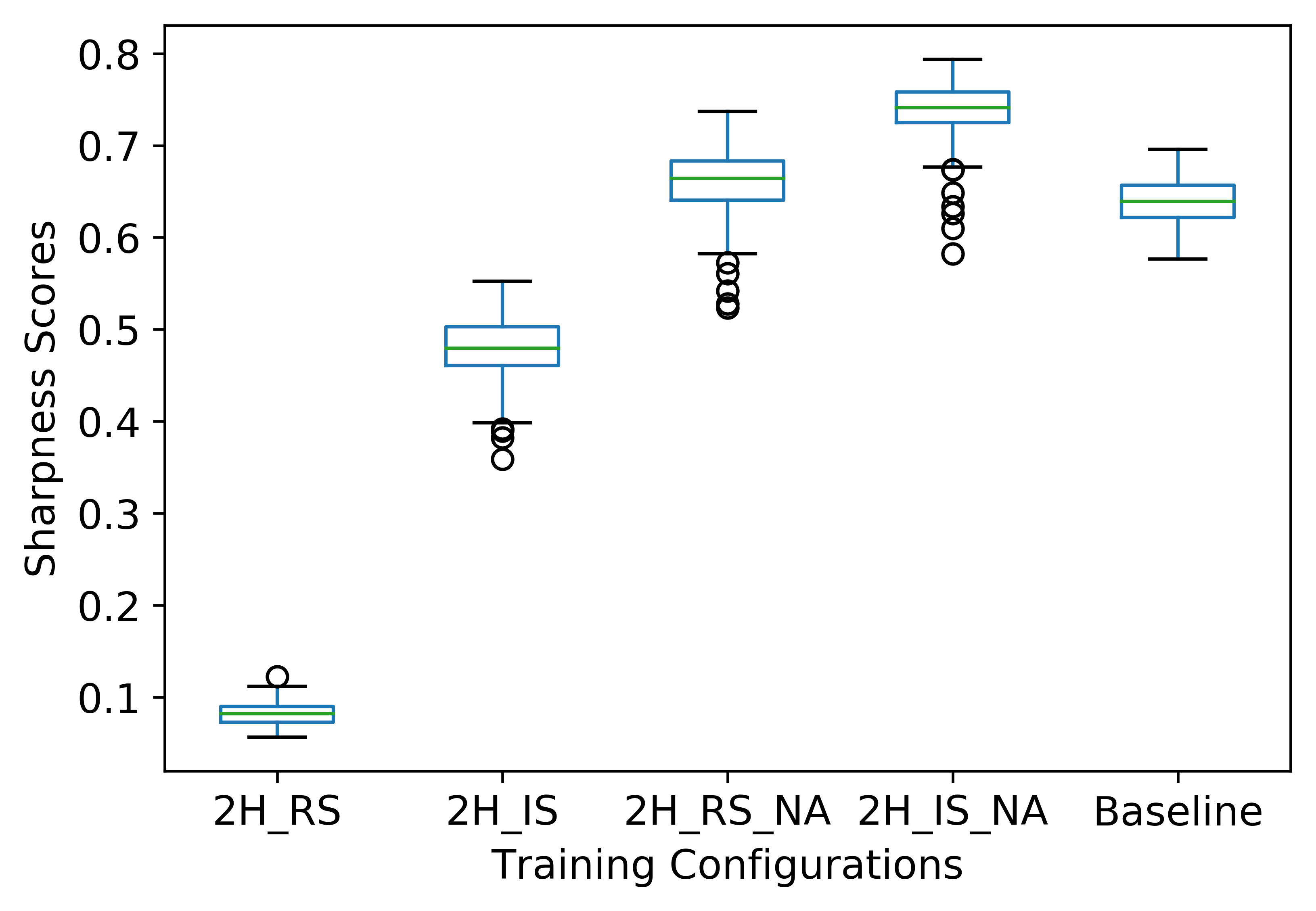}
  \caption{Sharpness scores of the learned attention weights across 174 prompts for the high-resource baseline and different low resource configurations. }
  \label{fig:alignment_sharp}
  \vspace{-0.5cm}
\end{figure}

We analyzed the learned attention weights across the input phoneme tokens and output mel spectrogram frame durations for each training configuration to verify our experiments' viability. In Fig.~\ref{fig:alignment}, weights are plotted for a representative speech, LJ002-0114.wav, from the LJ Speech dataset, for two configurations. We see a diagonal tendency in Fig.~\ref{fig:noise_align}, which indicates that the alignment between phonemes and mel frames has been learned with the 2H\_IS\_NA model. In contrast, Fig.~\ref{fig:align} shows that the attention weights are spread out flat, and no reasonable alignment is learned with the 2H\_RS model. 

Suppose the attention model learns a good alignment between the phonemes and the corresponding mel frames. In that case  the learned attention weight should be maximum (close to 1.0) at the aligned position and zero elsewhere. We define a sharpness score (inspired by \cite{schaefer_forwardtaco_20}) as the mean of the maximum attention weights across each mel frame. A sharpness score close to 1.0 indicates that the learned alignment is good. The boxplot of the sharpness scores of the learned attention weights for 174 sentences for different training configurations is plotted in Fig.~\ref{fig:alignment_sharp}. This figure shows that the random training data selection (RS) does not learn proper alignment, and the specific short duration set (IS) improves the alignment quite a bit. Both the noise augmented (NA) versions have learned good alignment, and the specific set with noise augmented training gets even sharper alignments than the baseline high resource training with the entire 23.5 hour database.   

\subsection{Quality Evaluation}
\label{subsec:sub-eval}

We used subjective Mean Opinion Score (MOS) tests and objective intelligibility tests based on Semantically Unpredictable Sentences (SUS)  for evaluation. These tests are described below.

The MOS tests were conducted using Absolute Category Rating as per P.808 \cite{rec2018p} using WebMUSHRA \cite{schoeffler2018webmushra} with 14 expert listeners from our laboratory having no reported hearing impairments, with an average age of 32.5~years. We used the first four lists from Harvard sentences \cite{Rothauser1969Harvard} with 40 sentences. At least 12 listeners rated each sentence in all the conditions.  

The SUS test was proposed as an objective measure of intelligibility for TTS models in \cite{benoit1996sus}. We synthesized all the 100 SUS texts which were part of the Blizzard challenge of 2005 \cite{black2005blizzard} with the TTS model under test. These were then transcribed with a pre-trained Speechbrain automatic speech recognition (ASR) \cite{speechbrain} recipe. The word error rate (WER) was measured using the Python JiWER package. A lower WER indicates better intelligibility. 

Table \ref{tab:list_test} gives the results from both tests, showing that the proposed model with 2 hours of specific training data and noise augmentation (2H\_IS\_NA) performs almost as well as the baseline version trained with 23.5 hours of data. The other configurations are progressively worse. Further, we can also see that the model trained using a random low resource set without noise augmentation (2H\_RS) cannot synthesize any intelligible speech as the WERs here are higher than 100 percent.

We also evaluated the models with long sentences, e.g., the 46-word sentence from `There There' by Tommy Orange \cite{henneke_21}. The output for this sentence with the proposed low-resource approach and the baseline approach was similar in terms of intelligibility during informal listening. Both had WER of 9\% with the ASR test. 

\begin{table}[!t]
  \caption{Evaluation metrics across the different simulated low resource configurations. MOS ratings are listed with 95\% confidence intervals. The ASR based SUS WER are given in percent. }
  \label{tab:list_test}
  \centering
  \begin{tabular}{ l  c c}
    \toprule
    \multicolumn{1}{l}{\textbf{LJ Speech Set}} & 
    \multicolumn{0}{c}{\textbf{MOS ($\uparrow$)}} &
    \multicolumn{0}{c}{\textbf{SUS-WER ($\downarrow$)}} \\
    \midrule
    2H\_RS                     & $1.21 \pm 0.04$ & 133.1  \\
    2H\_RS\_NA                 & $2.17 \pm 0.09$ & 76.5  \\
    2H\_IS                     & $3.02 \pm 0.09$ & 38.2  \\
    2H\_IS\_NA  			   & $\boldsymbol{3.98 \pm 0.09}$ & 19.6  \\
    Baseline (23.5H)             & $3.97 \pm 0.09$ & \textbf{18.5}  \\
    \bottomrule
  \end{tabular}
  \vspace{-1em}
\end{table}

%% file: discuss.tex
\pdfoutput=1
\section{Conclusion}
\label{sec:conc}

We implemented a low-resource single-speaker TTS model with minor modifications to the Tacotron-2 architecture. We proposed a training approach that does not require a separate pre-trained model and does not suffer from the accent and style transfer issues commonly present in multi-speaker, multi-language approaches. The proposed approach uses only 2 hours of specific data and noise augmentation for training. Further, augmentation identifiers are used to learn augmentation embeddings. Using specific training data and noise augmentation improved the learned text and speech alignment and thereby the synthesis. We demonstrated that the quality of speech synthesized by the proposed approach trained with only 2 hours of specific data is comparable to the speech synthesized by the baseline architecture trained with 23.5 hours of data using subjective tests. Further, objective intelligibility tests were conducted that support this finding. These insights can be used to train TTS models using much less data and still achieve natural quality speech synthesis. They might be used to train TTS models for dialects and languages for which little data is available. In the future, we plan to improve the specification process to account for phoneme distribution balance and study the effects of the current approach on the prosody of sentences.  

%% file: ack.tex
\pdfoutput=1
\section{Acknowledgements}
\label{sec:ack}

Parts of this work have been supported by the SPEAKER project (FKZ 01MK20011A), funded by the German Federal Ministry for Economic Affairs and Climate Action. In addition, this work was supported by the Free State of Bavaria in the DSAI project. 

\balance

%% file: lastone.tex
\pdfoutput=1
\balance